# Investigation on different physical aspects such as structural, elastic, mechanical, optical properties and Debye temperature of $Fe_2ScM$ (M = P and As) semiconductors: a DFT based first principles study

Md. Lokman Ali[1*]

*Department of Physics, Pabna University of Science and Technology,
Pabna-6600, Bangladesh*
*lokman.cu12@gmail.com*
[*]Corresponding author.

Md. Zahidur Rahaman[2]

*Department of Physics, Pabna University of Science and Technology,
Pabna-6600, Bangladesh*
*zahidur.physics@gmail.com*

**Abstract**

With the help of first principles calculation method based on the density functional theory we have investigated the structural, elastic, mechanical properties and Debye temperature of $Fe_2ScM$ (M = P and As) compounds under pressure up to 60 GPa. The optical properties have been investigated under zero pressure. Our calculated optimized structural parameters of both the compounds are in good agreement with the other theoretical results. The calculated elastic constants show that $Fe_2ScM$ (M = P and As) compounds are mechanically stable up to 60 GPa. From the elastic constants, the bulk modulus *B*, shear modulus *G*, Young's modulus *E*, Poisson's ratio $\nu$ and anisotropy factor *A* are calculated by using the Voigt-Reuss-Hill approximation. The sound velocities and Debye temperature are also investigated from the elastic constants. The detailed analysis of all optical functions reveals that both the compounds are good dielectric material and for both phases similar reflectivity spectra have been observed showing promise as a good coating materials. All these calculations have been carried out by using CASTEP computer code.

## 1. Introduction

Heusler alloys [1] have attracted great interest among the research community during the last century due to their remarkable physical properties. The full Heusler alloys possess many attractive behaviour such as magnetic phenomena like itinerant and localized magnetism, antiferromagnetism, helimagnetism, Pauli paramagnetism or heavy ferionic behaviour [2,3,4,5]. Recently their prominent application as spintronics [6] and as shape memory alloys have been intensively discussed [7]. The first full Heusler alloys were of the form $X_2YZ$ and crystallize in the $L2_1$ structure [8]. A number of theoretical works have been carried out on the electronic and thermodynamic properties of $Fe_2ScP$, $Fe_2ScAs$ and $Fe_2ScSb$ compounds. Recently, S. Sharma and S.k. Pandey [9] have performed the electronic and transport properties of $Fe_2ScX$ (X = P, As and Sb) compounds by using full-potential linearized augmented plane wave method based on density functional theory along with Boltzmann transport theory.

However, to the best of our knowledge, the structural, elastic, mechanical and optical properties and Debye temperature of $Fe_2ScP$ and $Fe_2ScAs$ compounds with $L2_1$ structure are still unexplored. In addition, the effects of pressure on the structural, elastic and mechanical properties and Debye temperature of $Fe_2ScP$ and $Fe_2ScAs$ compounds have not been reported up to now. It is well known that pressure plays an important role on the physical properties of materials, such as structural, elastic and mechanical properties [10-12]. It is reported that high pressure contributes to the phase transition and change in physical and chemical properties of a material [13-15]. So the study of pressure effects on these compounds is necessary and significant.

The main goal of our present work is to run a thorough investigation on the structural, elastic, mechanical properties and Debye temperature of $Fe_2ScM$ (M = P and As) compounds under hydrostatic pressure from 0 to 60 GPa with a step of 20 GPa by using the plane wave pseudo potential density functional theory (DFT) method. Furthermore, the optical properties under zero pressure are also calculated and discussed. The remaining parts of this paper are organized as follows. In section 2, we briefly outline the computational method. The result and discussion are presented in section 3. Finally, a brief conclusion is drawn in section 4.

## 2. Computational methods

The investigations were carried out by using the Density Functional Theory (DFT) [16] based CASTEP computer code [17]. The exchange correlation potential was described by the generalized gradient approximation (GGA) [18] with the Perdew –Burke_Ernzerhof (PBE) [19] function. The k-point meshes for Brillouin zone were sampled using $8 \times 8 \times 8$ according to the Meskhorst-Pack scheme [20]. The plane wave basis set with an energy cut-off of 340 eV is used for all cases. The structural parameters of $Fe_2ScM$ (M = P and As) were determined by using the Broyden-Fletcher-Goldfarb-Shenno (BFGS) minimization technique [21]. This method usually gives the fast way of finding the lowest energy structure. In the geometry optimization, the criteria of convergence were set as follows: (i) $1.0 \times 10^{-5}$ eV/atom for the total energy (ii) 0.03eV/Å for the maximum force (iii) 0.05 GPa for the maximum stress and (iv) the maximum ionic displacement was 0.001Å.

The elastic stiffness constants of $Fe_2ScM$ (M = P and As) compounds were determined from first principles calculations by applying a strain with a finite value and calculating the resulting stress according to Hook's law [22]. In this case the criteria of convergence were set as follows: (i) $2.0 \times 10^{-6}$ eV/atom for energy (ii) 0.006 eV/Å for the maximum ionic force and (iii) maximum ionic displacement was set to $2.0 \times 10^{-4}$ Å. The maximum strain amplitude was set to be 0.003 in all cases.

## 3. Results and Discussion

### *3.1. Structural properties*

The atomic structure of $Fe_2ScM$ (M = P, As) compounds are known to crystallize in a cubic lattice of $L2_1$ type structure which has space group *Fm-3m* (225). The optimized crystal structure of $Fe_2ScM$ is shown in Fig. 1. In this figure, the Fe atoms occupy the 8c wyckoff position with the fractional coordinate (1/4, 1/4, 1/4), Sc atoms are placed on the 4a wyckoff position with fractional coordinates (0, 0, 0) and M atoms (i.e., M = P and As) are located at the 4b wyckoff position with (1/2, 1/2, 1/2) fractional coordinate. The equilibrium lattice parameter has a value of 5.705 Å and 5.850 Å for $Fe_2ScP$ and $Fe_2ScAs$, respectively [9]. The calculated lattice parameters are listed in Table 1. One can see from Table 1 that the calculated lattice parameters of our present work are 5.662 and 5.811 Å for $Fe_2ScP$ and $Fe_2ScAs$ respectively which exhibit 0.75 and 0.66 % deviation from the other theoretical results. Our calculated equilibrium lattice parameter $a_0$ using GGA approximation is in good agreement with the other theoretical values.

**Table 1.** Structural parameters of $Fe_2ScM$ (M = P and As) compounds under zero pressure, including lattice parameters $a_0$ (Å), volume $V_0$ (Å$^3$), bulk modulus $B_0$ (GPa) and its pressure derivative $dB/dP$.

| Materials | Referance | $a_0$ (Å) | $B$(GPa) | $dB/dP$ | $V_0$ (Å$^3$) | Deviation from other results (%) |
|---|---|---|---|---|---|---|
| $Fe_2ScP$ | Present | 5.705 | 187.64 | 5.189 | 185.68 | |
| | Theoretical [9] | 5.662 | --- | --- | 181.51 | 0.75 |
| $Fe_2ScAs$ | Present | 5.850 | 177.03 | 5.853 | 200.20 | 0.66 |
| | Theoretical [9] | 5.811 | --- | --- | 196.22 | |

For investing the influence of applied external pressure on the crystal structure of $Fe_2ScM$ (M = P and As) compounds we have investigated the variations of the lattice constants and unit cell volume of $Fe_2ScM$ (M = P and As) compounds under different pressure.

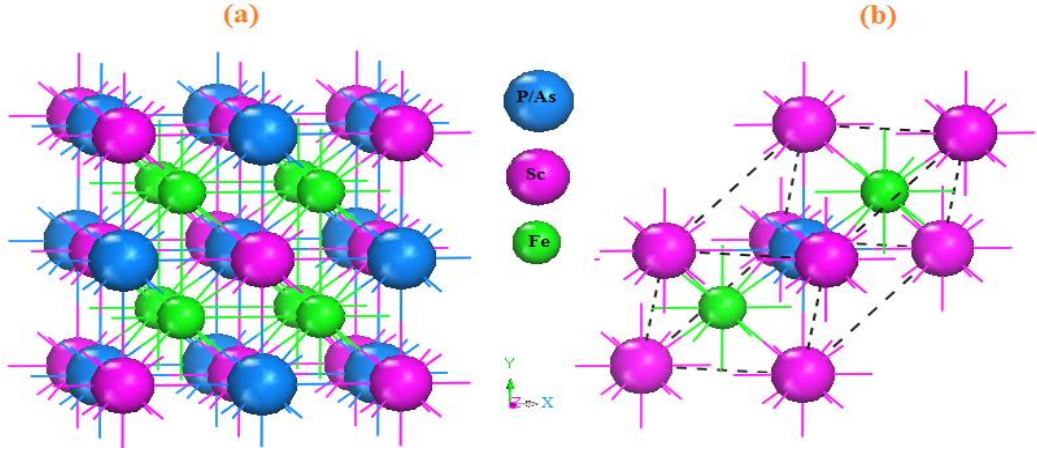

**Fig. 1.** The crystal structures of $Fe_2ScM$ (M = P and As) (a) conventional unit cell and (b) primitive cell.

The calculated structural parameters at different external pressure in the range from 0 GPa to 60 GPa with a step of 20 GPa are shown in Table 2. The relative changes of lattice parameters and unit cell volume as a function of external pressure of $Fe_2ScM$ (M = P and As) compounds are plotted in Fig. 2. It can be seen that the ratio of the $a/a_0$ decreases with the increase of pressure, where $a_0$ is the equilibrium lattice parameter at zero pressure. However, with the increase of pressure, the distance between atoms is reduced, which leads to the difficulty of compression of the crystal. The calculated unit cell volume at applied hydrostatic pressure in the range from 0 GPa to 60 GPa with a step of 20 GPa were used to construct the equation of state (EOS), which was fitted to a third order Brich-Murnaghan equation [23]. We obtained by least squares fitting the bulk modulus $B_0$ and its pressure derivative $dB/dP$. These are listed in Table 1. There is no experimental data available for the comparison.

**Table 2.** The calculated Lattice constant "$a$" and unit cell volume "$V$" of $Fe_2ScP$ and $Fe_2ScAs$ compounds under hydrostatic pressure.

| Materials | Pressure (GPa) | $a$ (Å) | $V$ (Å$^3$) |
|---|---|---|---|
| $Fe_2ScP$ | 20 | 5.546 | 170.58 |
|  | 40 | 5.444 | 161.34 |
|  | 60 | 5.357 | 153.73 |
| $Fe_2ScAs$ | 20 | 5.683 | 183.54 |
|  | 40 | 5.565 | 172.34 |
|  | 60 | 5.469 | 163.57 |

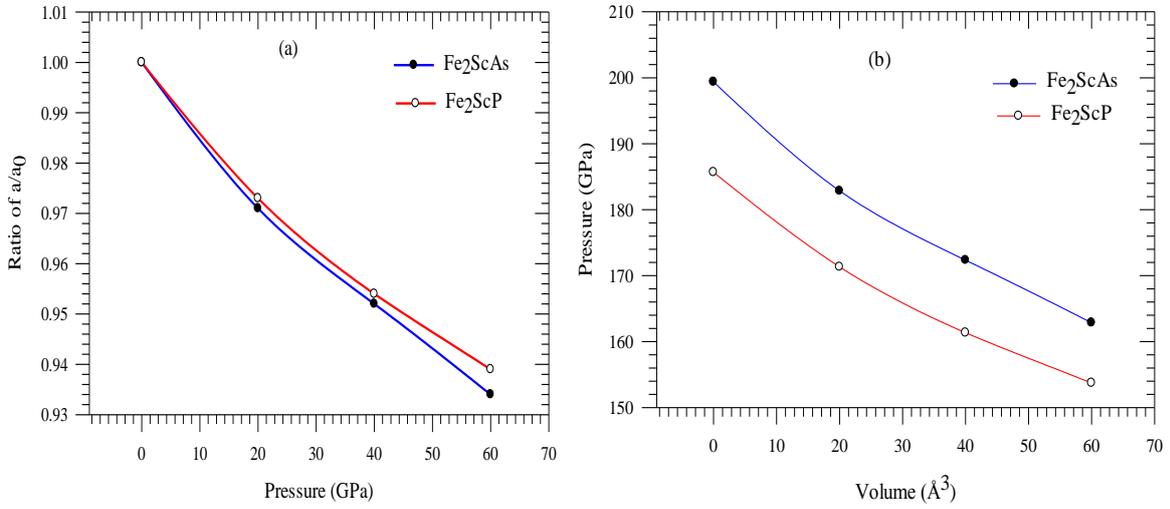

**Fig. 2.** (a) Variation of lattice parameters as a function of pressure and (b) Birch-Murnaghan equation of state for $Fe_2ScM$ (M = P and As)

### 3.2. Elastic constants

Elastic constants are very crucial parameter of the materials. It is used to determine the response of the crystal to applied external forces, which give the significant information concerning the nature of the forces operating in the crystal [24]. The elastic constants can also provide important information such as stability, brittleness, ductility, stiffness, bulk modulus, Young's modulus, shear modulus, Poisson's ratio and isotropic factor of materials. The elastic constants were determined from a linear fit of the calculated stress-strain function [22]. It is well known that a cubic crystal has three independent elastic constants $C_{11}$, $C_{12}$ and $C_{44}$. In this work, the calculated elastic constants of cubic $Fe_2ScM$ (M = P, As) semiconductors at zero pressure are presented in Table 3.

For cubic crystal, the elastic constants need to satisfy the following mechanical stability criteria [25],

$$C_{11} > 0, \ C_{44} > 0, \ C_{11} - C_{12} > 0 \ \text{and} \ C_{11} + 2C_{12} > 0 \qquad (1)$$

All of the elastic constants of $Fe_2ScM$ (M = P, As) semiconductors satisfy the above criteria of mechanical stability indicating that the structure of $Fe_2ScM$ (M = P, As) compounds are mechanically stable. In Table 3 we listed the calculated elastic constants of $Fe_2ScM$ (M = P, As) compounds at fixed values of the applied hydrostatic pressure in the range of 0-60 GPa with a step of 20 GPa. To the best of our knowledge, there are no experimental and other theoretical data in literature for the elastic constants of $Fe_2ScM$ (M = P, As) compounds for comparison. So we consider our obtained results as prediction study which still awaits an experimental confirmation in future.

We also report the pressure dependence behavior of elastic constants of $Fe_2ScM$ semiconductors as shown in figure 3. It can be seen that the $C_{11}$, $C_{12}$ and $C_{44}$ increases with the increase of pressure, among which $C_{11}$ is more sensitive than $C_{12}$ and $C_{44}$ to the change of pressure. A longitudinal strain causes a change in $C_{11}$, where $C_{11}$ represents the elasticity in length, Whereas $C_{12}$ and $C_{44}$ are shear constant and represents the elasticity in shape [26]. A transverse strain normally could cause a change in shape, but no change in volume [27].

**Table 3.** Calculated values of the elastic constants $C_{ij}$ (GPa) and the Cauchy pressure ($C_{12}$-$C_{44}$) of $Fe_2ScP$ and $Fe_2ScAs$ compounds at P = 0 GPa.

| *Materials* | *Pressure (GPa)* | $C_{11}$ | $C_{12}$ | $C_{44}$ | $C_{12}$ - $C_{44}$ |
|---|---|---|---|---|---|
| **$Fe_2ScP$** | 00 | 391.59 | 85.67 | 79.96 | 5.710 |
| | 20 | 550.35 | 161.96 | 122.95 | 39.01 |
| | 40 | 681.02 | 220.89 | 156.22 | 64.67 |
| | 60 | 802.25 | 277.35 | 185.45 | 91.90 |
| **$Fe_2ScAs$** | 00 | 350.16 | 90.46 | 88.97 | 1.460 |
| | 20 | 517.21 | 182.53 | 137.97 | 44.56 |
| | 40 | 636.44 | 236.11 | 174.73 | 61.38 |
| | 60 | 751.96 | 296.19 | 207.01 | 89.18 |

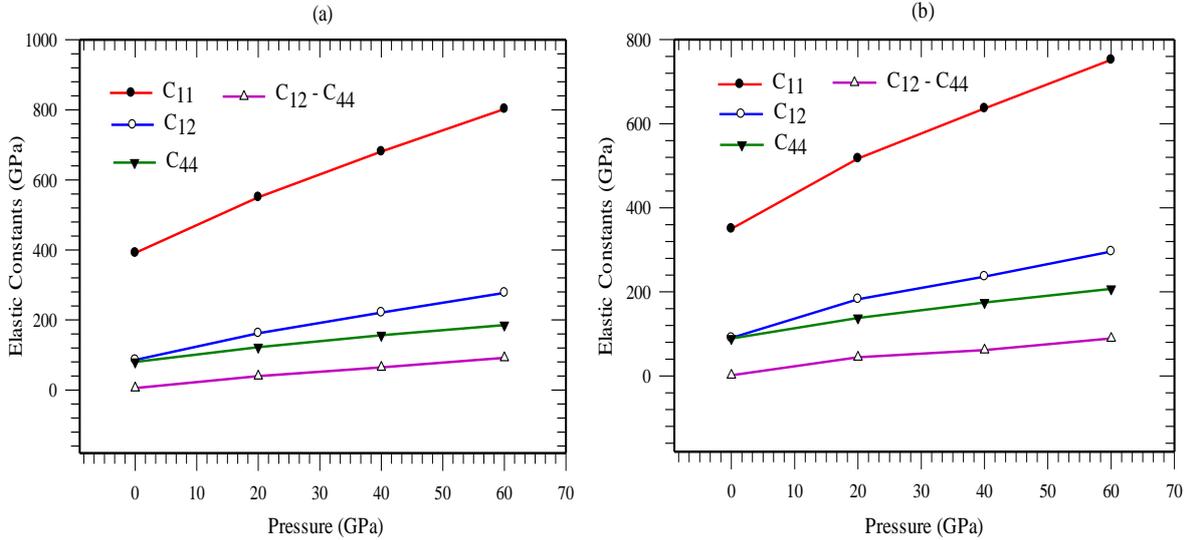

**Fig. 3.** Pressure dependence of elastic constants $C_{ij}$ ($C_{11}$, $C_{12}$, and $C_{44}$) and Cauchy pressure ($C_{12}$-$C_{44}$) of (a) $Fe_2ScP$ and (b) $Fe_2ScAs$ compounds.

## 3.3 Mechanical Properties

The most important mechanical properties of materials, such as the bulk modulus (*B*), shear modulus (*G*), Young's modulus (*E*), Poisson's ratio (*v*) and anisotropic factor (*A*), can be obtained from the calculated elastic constants using the Voigt-Reuss-Hill averaging scheme [28]. The Voigt-Reuss-Hill approximation method provides the values of the bulk and shear moduli. For the cubic crystal, the Voigt bound [29] of the bulk modulus $B_v$ and shear modulus $G_v$ are:

$$B_v = B_R = \frac{(C_{11} + 2C_{12})}{3} \quad (2)$$

$$G_v = \frac{(C_{11} - C_{12} + 3C_{44})}{5} \quad (3)$$

The Reuss bounds [30] of the bulk and shear moduli are:

$$B_v = G_v \quad (4)$$

$$\text{and,} \quad G_R = \frac{5C_{44}(C_{11} - C_{12})}{[4C_{44} + 3(C_{11} - C_{12})]} \quad (5)$$

According to the Hill approximation [28] method, the arithmetic average scheme of Voigt and Reuss gives the elastic moduli are expressed as following:

$$B = \frac{1}{2}(B_R + B_v) \quad (6)$$

$$G = \frac{1}{2}(G_v + G_R) \quad (7)$$

By using Hill's elastic moduli (*B*) and (*G*), the Young's modulus *(E)* and Poisson's ratio (*v*) can be directly calculated by the following equations:

$$E = \frac{9GB}{3B + G} \quad (8)$$

$$v = \frac{3B - 2G}{2(3B + G)} \quad (9)$$

The anisotropic factor (*A*) can be evaluated by using the following relation:

$$A = \frac{2C_{44}}{(C_{11} - C_{12})} \quad (10)$$

The calculated bulk modulus *B*, shear modulus *G*, Young's modulus *E*, Poisson's ratio *v* and anisotropy *A* of Fe$_2$ScM (M = P, As) semiconductors are listed in Table 4. The bulk modulus *B* can measure the resistance of materials to volume change by the applied pressure [31] and the shear modulus represents the resistance to plastic deformations upon shear stress [32]. The effects of pressure on the bulk modulus *B*, shear modulus *G* and Young's modulus *E* are presented in Fig. 4. It can be seen that all the parameters behave with almost linear relationships with the applied external pressure. As shown in Fig. 4, the values of bulk modulus *B* and shear modulus *G* at zero pressure are 187.64 and 103.99 GPa respectively for Fe$_2$ScP and 177.03 and 93.37 GPa for Fe$_2$ScAs. When the pressure reaches at 60 GPa, the values of *B* and *G* are 452.31 and 213.17 GPa for $Fe_2ScP$ and 448.11and 214.88 GPa for $Fe_2ScAs$. The present calculated results show that the pressure has a larger influence on the bulk modulus than the shear modulus.

The Young's modulus *E* is an important parameter which is defined as the ratio of the tensile stress to the tensile strain [34]. It is used to provide a measure of the stiffness of the solid. The larger the value of the Young's modulus *E*, the stiffer is the material. From Table 4 one can see that the calculated value of *E* increases with pressure almost linearly indicating that the pressure has a significant effect on the stiffness of $Fe_2ScM$ compounds.

The ratio of bulk modulus to shear modulus *B/G* is used to describe the plastic properties of materials. According to the Pugh criterion [34], a high value of *B/G* is associated with ductility, whereas a low value corresponds to the brittleness of the material. The critical value separating ductile and brittle material is 1.75. The calculated results are presented in Table 4. It can be seen that all the values of *B/G* are larger than the critical value 1.75, indicating that the $Fe_2ScM$ materials behave in a ductile manner.

**Table 4.** Calculated values of the bulk modulus (*B*), Shear modulus (*G*), Young's modulus (*E*), *B/G*, Poisson's ratio (*v*), and anisotropy factor (*A*) under various pressure for $Fe_2ScM$ (M = P and As) compounds.

| *Materials* | *Pressure (GPa)* | *B* | *G* | *E* | *B/G* | *v* | *A* |
|---|---|---|---|---|---|---|---|
| **$Fe_2ScP$** | 00 | 187.64 | 103.99 | 263.32 | 1.80 | 0.26 | 0.52 |
| | 20 | 291.42 | 147.34 | 378.26 | 1.97 | 0.28 | 0.80 |
| | 40 | 374.25 | 182.49 | 470.92 | 2.05 | 0.29 | 0.67 |
| | 60 | 452.31 | 213.17 | 552.68 | 2.12 | 0.30 | 0.70 |
| **$Fe_2ScAs$** | 00 | 177.03 | 93.37 | 238.22 | 1.89 | 0.27 | 0.68` |
| | 20 | 294.09 | 149.05 | 382.52 | 1.97 | 0.28 | 0.82 |
| | 40 | 369.55 | 184.49 | 474.51 | 2.00 | 0.285 | 0.87 |
| | 60 | 448.11 | 214.88 | 555.80 | 2.08 | 0.29 | 0.90 |

Fig. 4 shows the effect of pressure on the value of *B/G*. It can be seen that the value of *B/G* increases with increasing pressure, which indicates that pressure can improve the ductility of $Fe_2ScM$ compounds.

In addition, Nye [35] reported that the upper and lower values of Poisson's ratio *v* for central force solids are 0.50 and 0.25, respectively. In our present work, the calculated Poisson's ratio *v* of $Fe_2ScM$ compounds at zero pressure is larger than the lower limit value of 0.25, which indicates that the interatomic forces in $Fe_2ScM$ are central forces. As shown in Fig. 5, the obtained value of the Poisson's ratio *v* increases with the increase of pressure, this indicates that the interatomic forces of $Fe_2ScM$ are central forces.

The anisotropy factor is an important indicator of the degree of anisotropy in the solid structure [36]. When the value *A* = 1 means a completely isotropic material. As shown in Table 4, that the calculated anisotropy factor *A* is less than 1 for both compounds, which indicating that both the compounds show elastically anisotropic nature. The effect of pressure on the values of *A* is also shown in Fig. 5. It can be seen that the obtained value of *A* for $Fe_2ScP$ and $Fe_2ScAs$ at zero pressure are 0.52 and 0.68 respectively and it increases with pressure monotonously, finally reaching the value 0.90 and 0.70 at 60 GPa, which means that the $Fe_2ScM$ gradually becomes isotropic with increasing pressure.

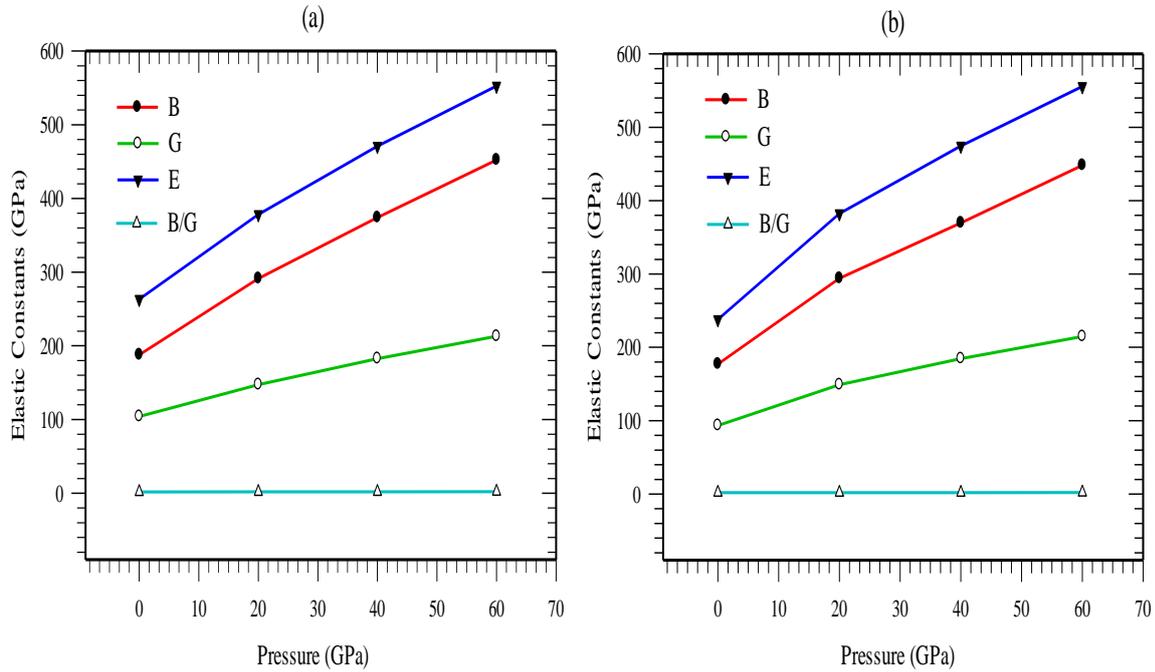

**Fig.4**. Variation of the bulk modulus *B*, shear modulus *G*, Young's modulus *E*, *B/G* of (a) $Fe_2ScP$ and (b) $Fe_2ScAs$ compounds under various pressure.

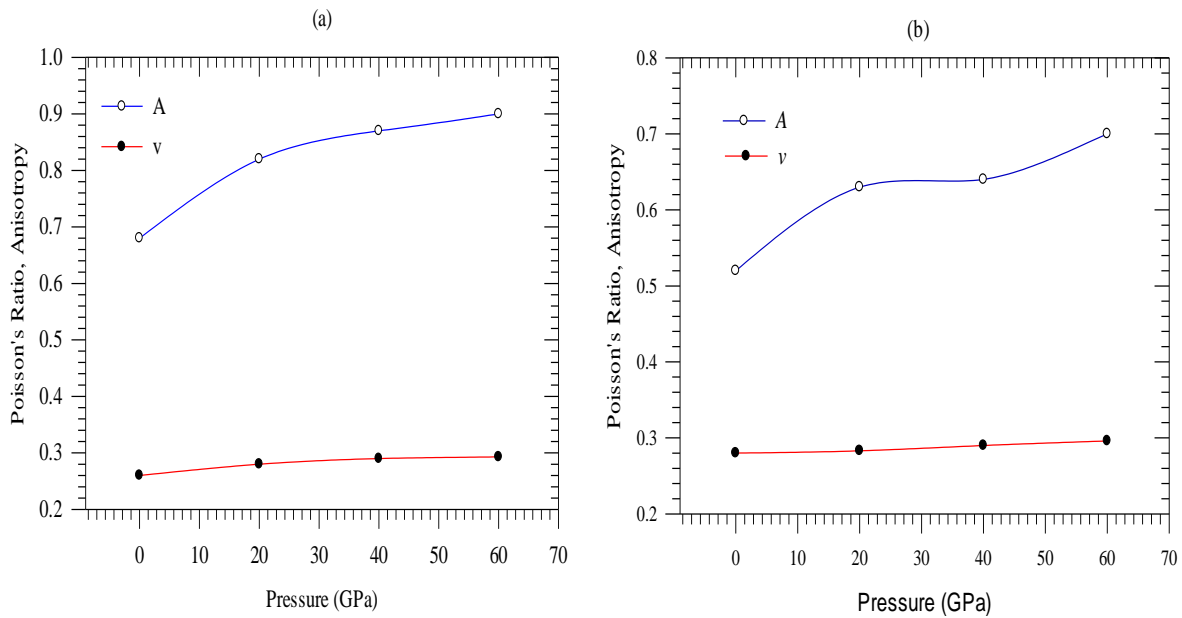

Fig. 5. The calculated Poisson's ratio and anisotropy factor of (a) $Fe_2ScP$ and (b) $Fe_2ScAs$ compounds at different pressure.

*3.4. Debye temperature*

The Debye temperature ($\Theta_D$) of materials is a crucial parameter plays an important role in many physical properties of solids, such as specific heat, elastic constants and melting temperature [37]. It is also directly connected with thermal vibration of atoms [32]. In addition, it reflects the structural stability and the strength of bonds of the solids. The Debye temperature of materials can be calculated from elastic constants. Using the average sound velocity $V_m$, the Debye temperature $\Theta_D$ of Fe$_2$ScM compounds was calculated by using the following equation [38]:

$$\theta_D = \frac{h}{k_B}\left(\frac{3N}{4\pi V}\right)^{\frac{1}{3}} \times v_m \qquad (11)$$

Where *h* is planck's constant, *K* is Boltzmann's constant, N is the number of atoms per formula unit and V is the atomic volume. The average wave velocity $V_m$ can be obtained by using the following equation [39]:

$$v_m = \left[\frac{1}{3}\left(\frac{2}{v_t^3} + \frac{1}{v_l^3}\right)\right]^{-\frac{1}{3}} \qquad (12)$$

Where $V_t$ and $V_l$ are the transeverse and longitudinal sound velocity respectively. The transverse $V_t$ and longitudinal $V_l$ sound velocity can be calculated using the shear modulus *G* and the bulk modulus *B* from Navier's equation [40]:

$$v_l = \left(\frac{3B + 4G}{3\rho}\right)^{\frac{1}{2}} \qquad (13)$$

And

$$v_t = \left(\frac{G}{\rho}\right)^{\frac{1}{2}} \qquad (14)$$

Where $\rho$ (= M/V) is the density, *M* is the molecular weight. The calculated longitudinal sound velocities $V_l$, transverse sound velocities $V_t$ and Debye temperature $\Theta_D$ as well as the density for Fe$_2$ScM compounds with a step of 20 GPa from 0 GPa to 60 GPa are listed in Table 5. The pressure dependence of Debye temperature of Fe$_2$ScM compounds is shown in Fig. 6. We can see that the Debye temperature increases with increasing pressure. No experimental and theoretical data is available in literature about the Debye temperature of these compounds; therefore it is difficult to make a comparison. Thus the present results can be served as a prediction study for future investigations.

**Table 5**. Pressure dependence of density ρ (kg/m³), transverse sound velocity $v_t$ (m/s), Longitudinal sound velocity $v_l$ (m/s), average wave velocity $V_m$ (m/s), and Debye temperature $\Theta_D$ (K) for the Fe$_2$ScM (M = P and As) compounds.

| Materials | Pressure (GPa) | ρ(kg/m³) | $v_t$ | $v_l$ | $v_m$ | $\Theta_D$(K) |
|---|---|---|---|---|---|---|
| Fe$_2$ScP | 00 | 6702.70 | 3938.86 | 6977.16 | 4381.25 | 576.75 |
| | 20 | 7294.11 | 4494.42 | 8178.38 | 5009.91 | 678.42 |
| | 40 | 7701.86 | 4867.67 | 8954.58 | 5430.44 | 749.14 |
| | 60 | 8104.57 | 5128.59 | 9533.05 | 5725.88 | 802.72 |
| Fe$_2$ScAs | 00 | 7650.00 | 3493.59 | 6278.12 | 3890.50 | 499.45 |
| | 20 | 8360.60 | 4222.26 | 7677.60 | 4706.28 | 621.93 |
| | 40 | 8895.34 | 4554.12 | 8318.51 | 5077.93 | 685.28 |
| | 60 | 9386.50 | 4784.60 | 8846.64 | 5339.77 | 733.27 |

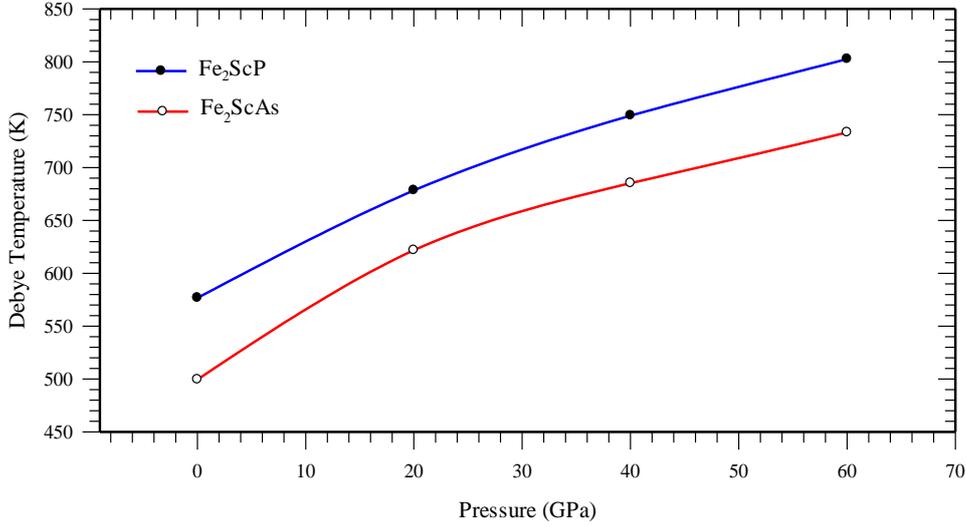

**Fig. 6.** Variation of Debye temperature $\Theta_D$ (K) of Fe$_2$ScP and Fe$_2$ScAs compounds at different pressure.

### *3.5. Optical properties*

The study of the optical properties of Fe$_2$ScM (M = P and As) with different photon energies is crucial for better understanding of the electronic structure of these compounds. The frequency dependent of complex dielectric function is defined as $\varepsilon(\omega) = \varepsilon_1(\omega) + i\varepsilon_2(\omega)$, where, $\varepsilon_1(\omega)$ and $\varepsilon_2(\omega)$ are the real and imaginary part of the dielectric functions, respectively. The imaginary part $\varepsilon_2(\omega)$ of the dielectric function can be expressed from the momentum matrix elements between the occupied and the unoccupied electronic states and can be calculated directly using [41]

$$\varepsilon_2(\omega) = \frac{2e^2\pi}{\Omega\varepsilon_0} \sum_{k,v,c} \left|\langle \psi_k^c | \hat{u} \cdot \vec{r} | \psi_k^v \rangle\right|^2 \delta(E_k^c - E_k^v - E) \tag{15}$$

Where $\omega$ is the frequency of light, e is the electronic charge, $\hat{u}$ is the vector defining the polarization of the incident electric field, and $\psi_k^c$ and $\psi_k^v$ are the conduction and valence band wave functions at $k$, respectively. The real part of the dielectric function is obtained from the imaginary part $\varepsilon_2(\omega)$ by using the Kramers-Kronig transform. The remaining optical functions, such as refractive index, absorption spectrum, loss function, reflectivity and conductivity are given by Eqs. (49) – (54) in Ref. [41]. Fig. 6 and Fig. 7 show the optical functions of $Fe_2ScM$ (M = P and As) compounds calculated for the photon energies up to 70 eV for polarization vectors [100] and [001]. We have used 0.5 eV Gaussain smearing for the present calculations.

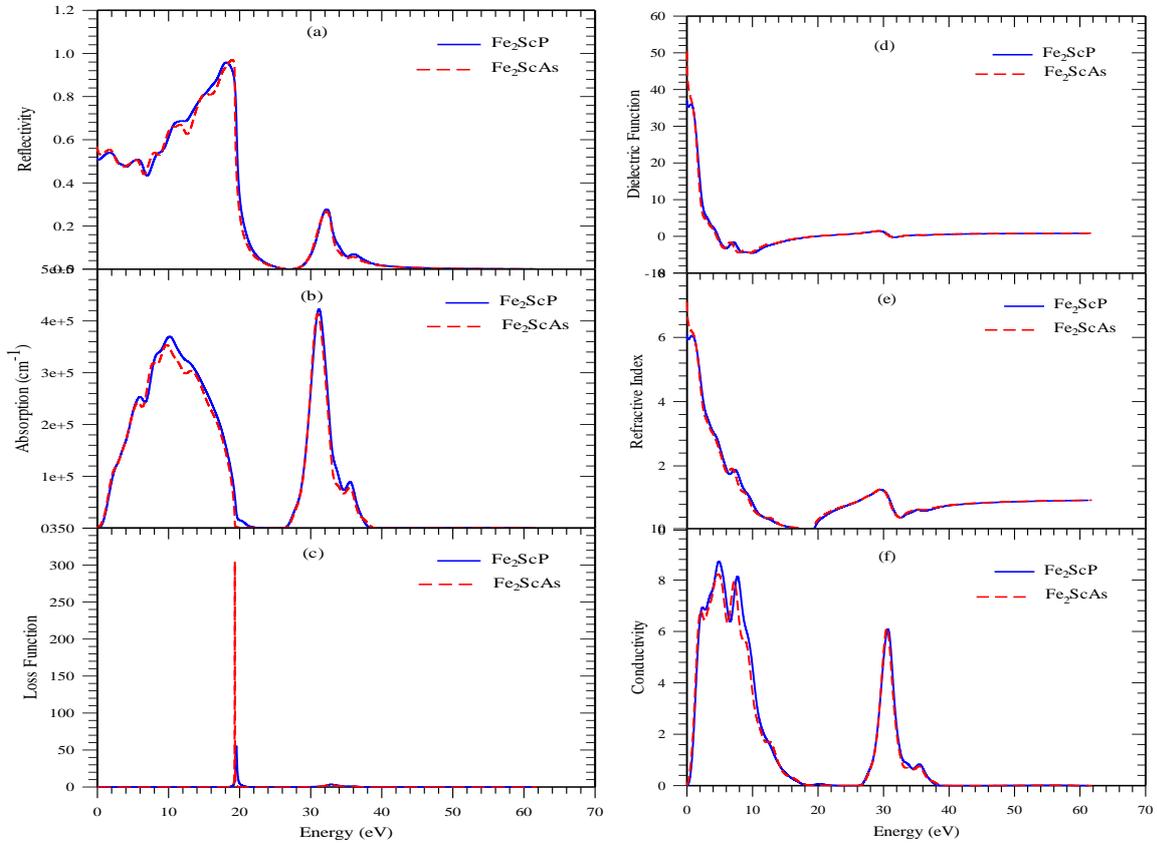

Fig. 6. The optical functions (a) reflectivity, (b) absorption, (c) Loss function, (d) dielectric function, (e) refractive index, (f) conductivity of $Fe_2ScM$ (M = P and As) for polarization vector [100].

Fig. 6 (a) shows the reflectivity spectra of $Fe_2ScM$ as a function of incident light energy. We noticed that the reflectivity of $Fe_2ScP$ and $Fe_2ScAs$ compounds having nearly similar characteristics starts with a value of ~0.52 – 0.54 decreases and then rise again to reach maximum value of ~0.96 – 0.27 between 18.5 eV and 32 eV. Thus both the phases possess roughly similar reflectivity spectra showing promise as good coating materials.

The absorption coefficient spectrum provides important information about optimum solar energy conversion efficiency and it indicates how far light energy (wavelength) can penetrate into the material before being absorbed [42]. The absorption spectra of $Fe_2ScM$ compounds as shown in Fig. 6

(b) reveal the semiconducting nature of these compounds, since the spectra starts from 0.03eV and 0.035 eV for $Fe_2ScP$ and $Fe_2ScAs$, respectively. It has two peaks at 9.9 eV and the other at 30 eV besides having a shoulder at lower energy. These compounds possess good absorption coefficient near 9.9 eV and 30 eV energy regions.

The electron energy loss function of materials is a key optical parameter to describe the energy loss of a fast electron passing through a material is usually large at the plasma frequency [43]. Fig. 6 (c) presents the energy loss function as a function of photon energy. It can be seen that the prominent peaks of both compounds at 19.8 eV, which indicates rapid reduction in the reflectance. This peak represents the feature that is associated with plasma resonance and the corresponding frequency is called bulk plasma frequency [44].

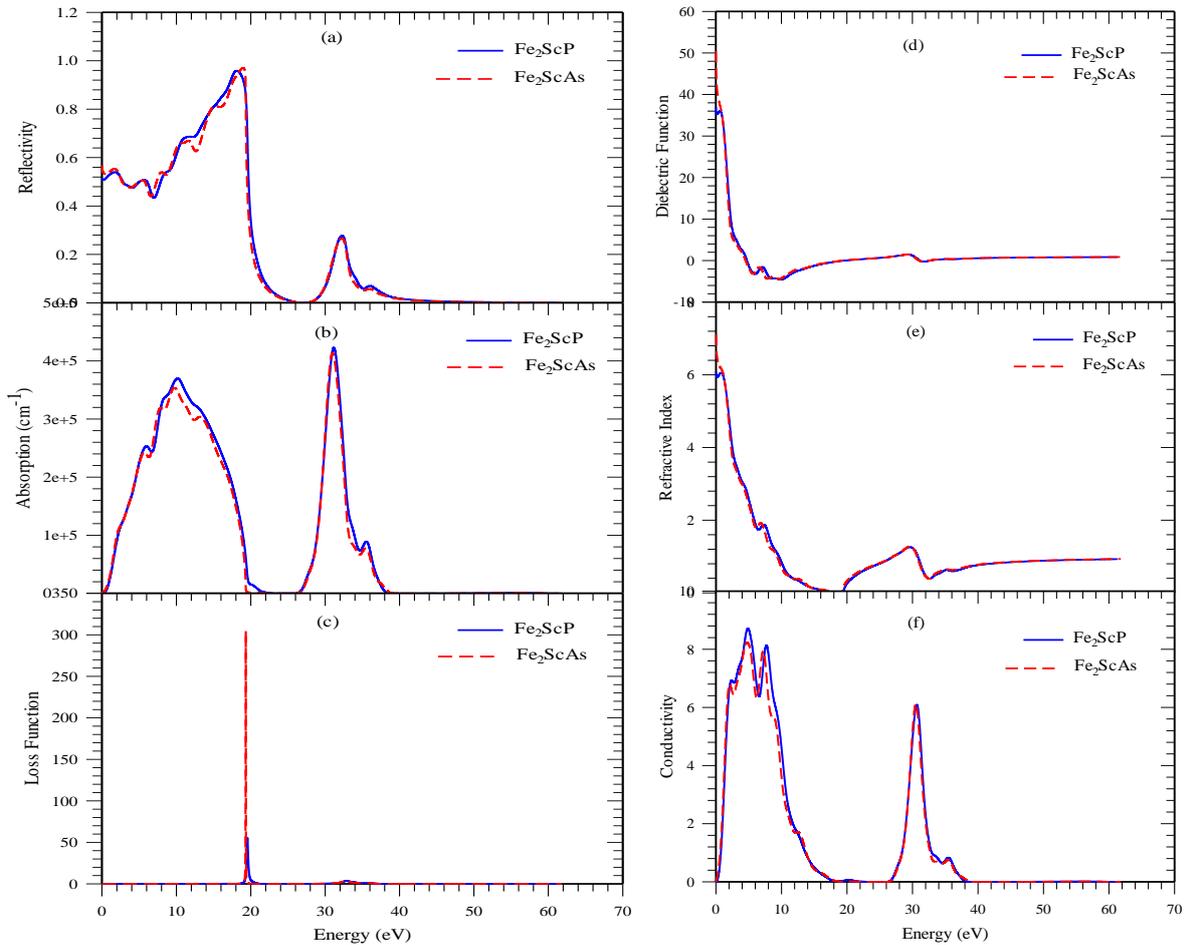

Fig. 7. The optical functions (a) reflectivity, (b) absorption, (c) loss function, (d) dielectric function (e) refractive index, (f) conductivity of $Fe_2ScM$ (M = P and As) for polarization vector [001].

The refractive index of an optical medium is a dimensionless number to describe how light or any other radiation propagates through the medium. The real parts of refractive indices of $Fe_2ScP$ and $Fe_2ScAs$ are displayed in Fig. 6 (e). The static refractive indices of $Fe_2ScP$ and $Fe_2ScAs$ for polarization direction [100] are 6 and 7.2, respectively. From Fig. 8 (e), it is noticed that the refractive indices of both compounds are higher in the low energy region and gradually decreased in the high energy region or ultraviolet region.

The conductivity spectrum of $Fe_2ScM$ compounds as a function of photon energy is shown in Fig 6 (f). We can see that the observed optical photoconductivity spectra have several maxima and minima within the energy range studied. Since both the materials have small band gap evident from band structure [9] the photoconductivity starts from ~ 0.03 eV as shown in Fig. 6 (f). Both of the compounds are electrically conductive when the incident radiation has energy within the range 3.5 eV-8 eV and 29 – 32 eV. There is no photoconductivity when the photon energy is higher than 39 eV.

The dielectric function is a fundamental optical parameter to describe the absorption and polarization properties of materials. The investigated dielectric constant of $Fe_2ScP$ and $Fe_2ScAs$ are shown in Fig. 6 (d) as a function of photon energy up to 70 eV. It can be seen that the dielectric constant becomes zero at about 9.54 eV for $Fe_2ScP$ and 9.55eV for $Fe_2ScAs$ indicating that both compounds become transparent above these energy respectively. The investigated result is in good consistent with the results obtained from the absorption spectra indicating the accuracy of our present DFT based calculation. The calculated static dielectric function of $Fe_2ScP$ is 36 and for $Fe_2ScAs$ the value is 50 indicating that $Fe_2ScAs$ is a good dielectric material than $Fe_2ScP$.

## *4. Conclusion*

In this paper, we have investigated the structural, elastic, mechanical, optical properties and Debye temperature of ternary compounds $Fe_2ScM$ (M = P and As) under various pressure up to 60 GPa using the first principle calculations based on density functional theory within the generalized gradient approximation. The main conclusions are as follows:

(a) The calculated lattice parameters are in good consistent with the previous theoretical results. The effect of pressure on lattice constants and cell volume are also investigated according to which the lattice constants and cell volume of the compounds decrease with the increase in pressure.
(b) Our obtained results of elastic constants exhibit that the structure of Fe2ScM (M = P and As) compounds are mechanically stable with the pressure up to 60 GPa. The pressure dependence of the elastic modulus (*B*, *G* and *Y*), *B/G*, Poisson's ratio *v* and anisotropy factor *A* have been also investigated. Using the Pugh's rule, the ductility of these compounds has been analyzed and the results show that both of the materials are ductile in nature.
(c) The pressure effects on Debye temperature $\Theta_D$ have been also investigated. The Debye temperature $\Theta_D$ of $Fe_2ScM$ (M = P and As) compounds increases with increasing pressure.
(d) The study of optical properties revels that $Fe_2ScM$ is a better dielectric material. It is also found that the reflectivity is high in visible-ultraviolet region (7 eV to 18.5 eV) for both $Fe_2ScP$ and $Fe_2ScAs$ compounds, showing promise as good coating materials.